\documentclass[aps,prl,twocolumn,superscriptaddress,nofootinbib,floatfix]{revtex4-2}

\pdfoutput=1

\usepackage{graphicx}
\usepackage{amsmath,amsfonts,amssymb}
\usepackage{xcolor}
\usepackage[normalem]{ulem}
\usepackage{hyperref}

\hypersetup{
  colorlinks = true,
  citecolor  = red,
  linkcolor  = blue,
  urlcolor   = blue
}

\definecolor{ao(english)}{rgb}{0.0, 0.5, 0.0}

\newcommand{\noi}{\noindent}
\newcommand{\beq}{\begin{equation}}
\newcommand{\eeq}{\end{equation}}
\newcommand{\bea}{\begin{eqnarray}}
\newcommand{\eea}{\end{eqnarray}}

\newcommand{\aend}{a_{\rm end}}

\newcommand{\Hend}{H_{\rm end}}

\graphicspath{{./figs/}}
\begin{document}


\title{Seesaw reheating}


\author{Yann Mambrini}
\email{yann.mambrini@ijclab.in2p3.fr}
\affiliation{
	Universit\'e Paris-Saclay, CNRS/IN2P3, IJCLab, 91405 Orsay, France 
}

\begin{abstract}
We introduce the Seesaw Reheating scenario, in which the inflaton transfers its energy to a long-lived intermediate scalar associated with the spontaneous breaking of lepton number before the Universe is reheated through its decay into right-handed neutrinos. As a result, the reheating temperature is no longer determined by the inflaton decay width but by the dynamics of the seesaw sector. We derive analytical solutions describing the complete reheating history, revealing two characteristic features of this scenario: the relativistic time dilation of the intermediate scalar, which suppresses its decay and delays the transfer of energy to the thermal bath, and its subsequent transition from a relativistic to a non-relativistic regime, introducing a new characteristic timescale in the thermal history. Together, these effects lead to simple analytical expressions for the reheating temperature. When the intermediate scalar is identified with the field responsible for the spontaneous breaking of lepton number, the same framework naturally connects the reheating temperature to the origin of neutrino masses and provides a well-motivated setting for sterile-neutrino dark matter.
\end{abstract}

\maketitle
\twocolumngrid


\section{Introduction} \label{sec: introduction}

The reheating temperature, $T_{\rm RH}$, is one of the key quantities connecting inflation to the subsequent thermal history of the Universe. It determines the onset of the radiation-dominated era and plays a central role in processes such as baryogenesis, dark matter production, and Big Bang nucleosynthesis. Over the past decade, considerable effort has been devoted to constraining $T_{\rm RH}$ from cosmological observations, establishing a direct connection between inflationary observables and the reheating epoch~\cite{Martin:2010kz,Dai:2014jja,Garcia:2020eof,Garcia:2021iag}.
 While the only robust lower bound remains the requirement of successful Big Bang nucleosynthesis, $T_{\rm RH}\gtrsim3\,{\rm MeV}$, stronger constraints can be obtained in specific frameworks relating reheating to dark matter production~\cite{Henrich:2024rux,Henrich:2025gsd,Henrich:2025sli,Bertou:2026osq}. Looking ahead, future CMB and large-scale structure surveys such as LiteBIRD and Euclid are expected to substantially improve our ability to probe the reheating history and its underlying microphysics, opening the possibility of constraining the interactions responsible for transferring the inflaton energy to the primordial plasma~\cite{Drewes:2022nhu,Martin:2010kz,Dai:2014jja,Cook:2015vqa}.

In most existing studies, the reheating temperature is assumed to be determined directly by the inflaton decay width. Imposing gauge invariance at the inflationary scale, this may occur through couplings of the inflaton to the Standard Model Higgs sector~\cite{Garcia:2021iag,Garcia:2020eof,Gross:2015cwa,Tenkanen:2016twd}, or through direct couplings to right-handed neutrinos in seesaw-inspired inflationary models~\cite{Asaka:1999jb,Lazarides:1991wu,Buchmuller:2005eh,Han:2024qbw,Cosme:2024ndc,Datta:2025wfh}. Although the microscopic realization differs, these scenarios share the common feature 
that inflaton decays or scatterings directly initiate the reheating process, so that the reheating temperature remains controlled by the inflaton decay rate, 
even when thermalization is delayed~\cite{Amin:2018eta,Mukaida:2015ria}
or reheating proceeds through purely gravitational couplings~
\cite{Mambrini:2021zpp,Bernal:2023nqg,Clery:2021bwz,Co:2022bgh}.
Under this assumption, the reheating temperature is naturally interpreted as a direct probe of the inflaton sector.

Is the reheating temperature necessarily a direct probe of the inflaton? We argue that the answer need not be affirmative. Reheating may 
naturally proceed through a long-lived intermediate state, so that the 
temperature of the thermal bath is controlled by the lifetime of this 
mediator rather than by the inflaton itself. As a concrete realization, we consider the scalar responsible for the spontaneous breaking of $U(1)_{B-L}$, whose decay into right-handed neutrinos completes the reheating 
process.

We argue that this apparently simple modification qualitatively changes the physics of reheating. The reheating temperature is no longer determined by the inflaton decay width, but {\it by the properties of the long-lived intermediate scalar}. When this scalar is identified with the field responsible for spontaneous lepton-number breaking, the same framework naturally connects the reheating temperature to the seesaw scale and provides a well-motivated setting for sterile-neutrino dark matter. This defines a new reheating paradigm, which we refer to as \emph{Seesaw Reheating}, in which the reheating temperature probes the mediator connecting inflation to the visible sector rather than the inflaton itself.

\section{The Seesaw reheating mechanism}

The inflaton is assumed to oscillate around the minimum of a quadratic 
potential after inflation, behaving as pressureless 
matter. Instead of transferring its energy directly to the particles forming the primordial radiation bath, we consider the minimal realization 
of Seesaw Reheating,\footnote{Generalizations to arbitrary inflaton potentials, $V(\phi)\propto\phi^k$, as well as alternative reheating interactions (e.g. $\phi^2S^2$ scatterings) and more general realizations of Seesaw Reheating will be presented elsewhere.} described by the minimal Lagrangian
\begin{equation}
\mathcal{L} \supset
-\frac12 m_\phi^2\phi^2
-\frac12 m_S^2S^2
-\frac{\mu}{2}\phi S^2
-\frac12 M_N\overline{N_R^c}N_R
-\frac{y_N}{2}S\,\overline{N_R^c}N_R \,,
\end{equation}
which induces the decay chain
\begin{equation}
\phi\rightarrow SS\rightarrow (N_RN_R)(N_RN_R) \,,
\end{equation}
where $N_R$ denotes a Majorana right-handed neutrino.
The corresponding decay widths are
\begin{equation}
\Gamma_\phi=
\frac{\mu^2}{32\pi m_\phi}
\sqrt{1-\frac{4m_S^2}{m_\phi^2}}\,,
\qquad
\Gamma_S=
\frac{y_N^2m_S}{32\pi}
\left(1-\frac{4M_N^2}{m_S^2}\right)^{3/2}\,.
\end{equation}
The intermediate scalar is produced highly relativistically through the
decay $\phi\rightarrow SS$. Since inflaton decays occur continuously, the
scalar population cannot be accurately described by a single physical
momentum redshifting from the end of inflation. Instead, particles
produced at different epochs populate different momentum shells, leading
to a time-dependent momentum distribution. The effective decay rate is
therefore determined by the average Lorentz factor of the scalar
population,
\beq
\Gamma_S^{\rm eff}
=
\frac{\Gamma_S}{\langle\gamma_S\rangle}\, ,
\eeq
where $\langle\gamma_S\rangle$ denotes the production-weighted average
Lorentz factor of the continuously produced scalar population.
A complete treatment based on
the evolution of the full phase-space distribution will be presented
elsewhere. 

The post-inflationary evolution is governed by the coupled Boltzmann equations
\beq
\left\{
\begin{aligned}
\dot{\rho}_\phi+3H\rho_\phi
&=-\Gamma_\phi\rho_\phi\, ,\\
\dot{\rho}_S+3H(1+w_S)\rho_S
&=\Gamma_\phi\rho_\phi
-\Gamma_S^{\rm eff}\rho_S\, ,\\
\dot{\rho}_R+4H\rho_R
&=\Gamma_S^{\rm eff}\rho_S\, ,
\end{aligned}
\right.
\label{Eq:boltzmann}
\eeq
together with the Friedmann equation
\beq
3M_P^2H^2
=
\rho_\phi+\rho_S+\rho_R\, ,
\eeq
where $M_P=1/\sqrt{8\pi G}\simeq2.4\times10^{18}\,\mathrm{GeV}$ is the
reduced Planck mass.

For simplicity, we assume that the right-handed neutrinos produced in the decay of $S$ 
thermalize promptly with the primordial plasma, i.e. that their thermalization rate 
satisfies $\Gamma_{\rm th}\gg H$. Their energy is therefore transferred to the 
radiation bath on a timescale much shorter than the Hubble time.\footnote{For the benchmark considered in this Letter, thermalization becomes efficient during the build-up of the radiation bath, thereby justifying the effective description adopted here. A more general treatment including finite thermalization 
rates and departures from instantaneous thermal equilibrium will be presented in a 
forthcoming companion paper.} Their energy density can therefore be identified with 
that of the radiation bath, $\rho_R$, throughout the reheating process.

Note also that the trilinear interaction $\mu \phi S^2$ induces a time-dependent
effective mass for the mediator through the oscillating inflaton
condensate, which can delay the kinematic opening of the
$\phi\rightarrow SS$ decay channel. A quantitative treatment of this
effect requires a self-consistent evolution of the condensate together
with the effective decay width \cite{Garcia:2020eof,Garcia:2020wiy,Barman:2025lvk}. Since the purpose of the present Letter
is to establish the Seesaw Reheating mechanism, we defer this analysis
to a forthcoming companion paper and focus here on the minimal scenario
in which the decay width of $S$ is described by its vacuum expression.

\begin{figure}[!htb]
   \begin{minipage}{0.45\textwidth}
     \centering
     \includegraphics[width=0.99\linewidth]{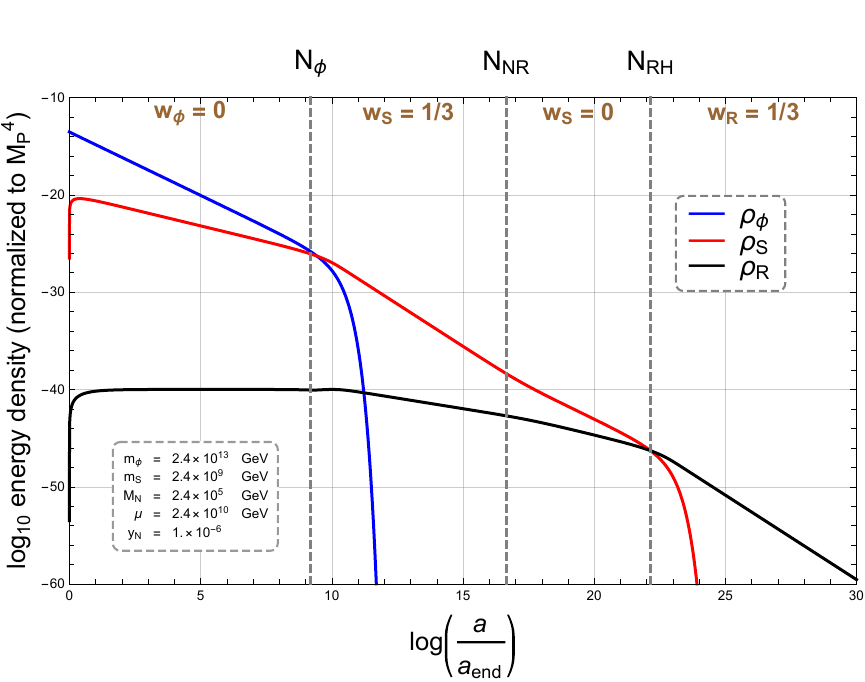}
     \caption{
Evolution of the post-inflationary energy budget in Seesaw Reheating.
The dashed lines mark the onset of efficient inflaton decay ($N_\phi$),
the relativistic-to-non-relativistic transition of the intermediate
scalar ($N_{\rm NR}$), and the onset of radiation domination
($N_{\rm RH}$).
}
 \label{Fig:seesawreheating}
   \end{minipage}
  
\end{figure}

Within this minimal framework, the intermediate scalar $S$ is produced
highly relativistically through the decay $\phi\rightarrow SS$. As the
scalar population redshifts after the end of inflaton decay, its equation
of state evolves continuously from $w_S\simeq1/3$ to $w_S\simeq0$,
introducing a distinct intermediate stage between inflaton domination and
radiation domination. As we show below, this transition leaves a
characteristic imprint on the reheating dynamics.
The evolution of $\rho_\phi$, $\rho_S$, and $\rho_R$, normalized to
$M_P^4$, is displayed in Fig.~\ref{Fig:seesawreheating} for the
representative benchmark shown in the figure, with
$\rho_\phi^{\rm end}=(10^{15}\,\mathrm{GeV})^4$.

We observe that during the inflaton-dominated era, the radiation energy
density increases only gradually. This distinctive behaviour originates
from the fact that the intermediate scalar is produced highly
relativistically, so that its decay into right-handed neutrinos is
initially suppressed by Lorentz time dilation. As the scalar redshifts,
its Lorentz factor decreases, thereby shortening its lifetime in the
cosmological frame. In other words, the cosmological expansion
progressively lifts the Lorentz suppression of the decay, so that the
transfer of energy to the thermal bath becomes increasingly efficient
and eventually overcomes Hubble dilution, causing $\rho_R$ to grow, as
observed in Fig.~\ref{Fig:seesawreheating}. Consequently, the plasma temperature increases only mildly throughout this stage, instead of exhibiting the large 
transient maximum temperature characteristic of conventional perturbative reheating.

Once
the inflaton has disappeared, the subsequent evolution is governed by the
intermediate scalar, whose equation of state evolves continuously from
$w_S\simeq1/3$ to $w_S\simeq0$ as it becomes non-relativistic. This
transition leaves a characteristic imprint on the reheating history before
the final transfer of energy to radiation through the decay of $S$ into
right-handed neutrinos. 

An important consequence of this setup is that the reheating temperature is no longer determined by the inflaton decay width, but instead by the lifetime of the intermediate seesaw scalar.

\section{Analytical Understanding of Seesaw Reheating}

Having established the main features of the Seesaw Reheating mechanism, we now turn to a more detailed analytical understanding of the evolution displayed in Fig.~\ref{Fig:seesawreheating}. In particular, we show that the characteristic changes of slope in the energy densities and the existence of a prolonged low-temperature stage follow from simple scaling arguments.

A first remarkable feature of Seesaw Reheating is the existence of an
extended low-temperature phase during which the radiation bath is built
up progressively. During the inflaton-dominated era,
\beq
\rho_\phi\simeq\rho_{\phi}^{\rm end}\left(\frac{\aend}{a}\right)^{3},
\qquad
H\simeq \Hend\left(\frac{\aend}{a}\right)^{3/2}\, .
\eeq

\noi
At early times, when $H\gg\Gamma_S^{\rm eff}$, the Boltzmann equation
(\ref{Eq:boltzmann}) reduces to
\beq
\frac{d\rho_S}{d\ln a}
+3(1+w_S)\rho_S
\simeq
\frac{\Gamma_\phi}{H}\rho_\phi\, ,
\eeq
where the source term scales as
$(\Gamma_\phi/H)\rho_\phi\propto a^{-3/2}$.
The source-dominated solution is therefore
\beq
\rho_S\propto a^{-3/2}\, ,
\eeq
in agreement with Refs.~\cite{Garcia:2020eof,Garcia:2021iag}.

On the other hand, the radiation bath obeys
\beq
\frac{d\rho_R}{d\ln a}
+4\rho_R
\simeq
\frac{\Gamma_S^{\rm eff}}{H}\rho_S\, ,
\eeq
where
$\Gamma_S^{\rm eff}=\Gamma_S/\langle\gamma_S\rangle$.
Continuous inflaton decays replenish the relativistic part of the scalar
population with newly produced particles. During the source-dominated
inflaton era, and neglecting the decay of $S$, the production-weighted
average energy is well approximated by
\beq
\langle E_S\rangle
\simeq
\frac35\,\frac{m_\phi}{2}\, ,
\eeq
following the treatment of continuous non-thermal production developed
in \cite{Garcia:2023dyf,Garcia:2026ulw}.
In the relativistic regime, this gives
\beq
\langle\gamma_S\rangle
\simeq
\frac35\,\frac{m_\phi}{2m_S}\, .
\eeq
Since $\rho_S/H$ also remains approximately constant during this period,
the radiation source term scales as
\beq
\frac{\Gamma_S^{\rm eff}}{H}\rho_S
\simeq
{\rm const.}\, ,
\eeq
leading to
\beq
\rho_R\simeq {\rm const.}\,,
\qquad
T\simeq {\rm const.}\, ,
\eeq
up to the decaying homogeneous contribution and corrections arising near
the end of inflaton domination. Seesaw Reheating therefore exhibits an
extended temperature plateau while the relativistic scalar population is
continuously replenished. Once inflaton decay becomes inefficient, the
remaining scalar distribution redshifts, the average Lorentz factor
decreases, and the transfer of energy to radiation becomes progressively
more efficient.

The energy transfer from the seesaw sector to radiation therefore remains
suppressed throughout most of the inflaton-dominated era because the
intermediate scalar population stays highly relativistic while being
continuously replenished by inflaton decays. As a result, the Universe
remains at a parametrically low temperature until the intermediate scalar
eventually becomes non-relativistic and efficiently reheats the plasma.
This naturally provides a framework for scenarios requiring a genuinely
low maximum temperature, such as the recently proposed stronger-coupling
freeze-in regime\footnote{Note that gravitational particle production may
constitute an irreducible source of radiation, potentially modifying the
phenomenology of specific low-temperature scenarios, as recently discussed
in Ref.~\cite{Gross:2026bhe}.}\cite{Cosme:2024ndc,Bertou:2026osq}.

A second remarkable feature of Seesaw Reheating is the
relativistic-to-non-relativistic transition of the mediator, which
introduces a new stage in the reheating history and accounts for the
second break in $\rho_S$ visible in
Fig.~\ref{Fig:seesawreheating}. Since inflaton decays continuously
populate the scalar sector, particles produced at different epochs become
non-relativistic at different times. The transition is therefore gradual
rather than instantaneous. We define the characteristic epoch
$N_{\rm NR}$ as the time at which the average energy of the scalar
population falls to twice its rest mass. For the benchmark shown in
Fig.~\ref{Fig:seesawreheating}, this transition occurs only after the
inflaton has disappeared, during the subsequent $S$-dominated era.

During the relativistic $S$-dominated stage, inflaton decays have ceased
and the scalar momentum distribution evolves only through cosmological
redshift. The average Lorentz factor consequently scales as
$\langle\gamma_S\rangle\propto a^{-1}$, implying
\beq
\rho_S\propto a^{-4}\,,
\qquad
H\propto a^{-2}\,,
\qquad
\Gamma_S^{\rm eff}
=
\frac{\Gamma_S}{\langle\gamma_S\rangle}
\propto a\, .
\eeq
The source term for radiation production therefore scales as
\beq
\frac{\Gamma_S^{\rm eff}}{H}\rho_S
\propto a^{-1}\, ,
\eeq
so that
\beq
\rho_R\propto a^{-1}\, ,
\eeq
up to the faster-decaying homogeneous solution. Once the scalar population
becomes non-relativistic, $\langle\gamma_S\rangle\simeq1$, while
\beq
\rho_S\propto a^{-3}\,,
\qquad
H\propto a^{-3/2}\, ,
\eeq
The radiation source term then scales as
\beq
\frac{\Gamma_S}{H}\rho_S
\propto a^{-3/2}\, ,
\eeq
implying
\beq
\rho_R\propto a^{-3/2}\, .
\eeq

The second change of slope visible in
Fig.~\ref{Fig:seesawreheating} therefore provides a direct signature of
the relativistic-to-non-relativistic transition of the intermediate
scalar. Unlike conventional reheating, where the radiation bath is
entirely governed by inflaton decay, Seesaw Reheating predicts an
additional characteristic timescale associated with the dynamics of the
mediator itself. 

The reheating temperature is defined by the onset of radiation domination,
\beq
\rho_R(a_{\rm RH})=\rho_S(a_{\rm RH})\,,
\eeq
from which
\beq
T_{\rm RH}
=
\left(
\frac{30\,\rho_R(a_{\rm RH})}
{\pi^2g_*(T_{\rm RH})}
\right)^{1/4}\,,
\eeq
where $g_*(T_{\rm RH})$ denotes the effective number of relativistic
degrees of freedom in the thermal bath at reheating; in the Standard
Model, $g_*=106.75$ at high temperature.
Despite the non-trivial relativistic evolution of the intermediate
scalar, the reheating temperature remains parametrically controlled by
its decay width,
\beq
\boxed{
T_{\rm RH}
\propto
\sqrt{\Gamma_S M_P}
\propto
y_N\sqrt{m_SM_P}\, ,
}
\eeq
up to corrections induced by the preceding relativistic evolution of the intermediate scalar.
This result replaces the standard perturbative reheating relation,
\beq
T_{\rm RH}^{\rm std}
\propto
\sqrt{\Gamma_\phi M_P}\,,
\eeq
demonstrating that the reheating temperature probes the seesaw sector rather than the inflaton.

Beyond the modified reheating dynamics, Seesaw Reheating also establishes
a direct connection between the reheating temperature and the origin of
neutrino masses. When $S$ is identified with the scalar responsible for
the spontaneous breaking of lepton number, acquiring the vacuum
expectation value $v_S$,
\beq
M_N=\frac{y_Nv_S}{\sqrt2},
\qquad
m_S=\sqrt{2\lambda_S}\,v_S\, ,
\eeq
so that
\beq
T_{\rm RH}
\propto
\lambda_S^{1/4}
\sqrt{y_NM_PM_N}\, .
\eeq
The reheating temperature is therefore directly connected to the seesaw
scale through the parameters governing Majorana mass generation. In the
minimal one-generation realization, the type-I seesaw relation
\beq
m_\nu\simeq
\frac{y_\nu^2v^2}{M_N}
\eeq
further relates the reheating temperature to the neutrino Yukawa
coupling. In realistic multi-generation seesaw models, this relation is
generalized through the full Yukawa matrix. Throughout this work, we
assume that the right-handed neutrinos produced in the decay of $S$
possess sufficiently large Yukawa interactions with the Standard Model to
thermalize rapidly after their production, allowing their energy density
to be identified with the radiation bath. A more complete treatment of
the thermalization dynamics of the different right-handed neutrino
species will be presented elsewhere.

In this framework, the reheating temperature is no longer an independent
parameter of the inflaton sector, but becomes linked to the physics
responsible for generating neutrino masses. Consequently, measurements
of the neutrino sector, combined with information on the scalar sector,
could provide direct information on the reheating temperature,
establishing an unexpected bridge between laboratory measurements of
neutrino properties and the thermal history of the early Universe.

\section{Discussion and conclusions}

The reheating temperature is commonly regarded as one of the few observables directly connected to the inflaton sector. This interpretation, however, relies on the implicit assumption that the inflaton transfers its energy directly to the primordial plasma. In this Letter, we have shown that this assumption need not hold. Whenever the inflaton first decays into a long-lived intermediate state, the reheating temperature is instead determined by the dynamics of this mediator.

In the minimal realization considered here, the mediator is identified with the scalar associated with the spontaneous breaking of lepton number, whose decay into right-handed neutrinos completes the reheating process. This defines the Seesaw Reheating paradigm, in which
\beq
T_{\rm RH}\propto\sqrt{\Gamma_SM_P}\,,
\eeq
replacing the standard scaling
\beq
T_{\rm RH}^{\rm std}\propto\sqrt{\Gamma_\phi M_P}\,.
\eeq
The reheating temperature therefore becomes a direct probe of the seesaw sector rather than of the inflaton itself.

Beyond shifting the reheating temperature from the inflaton sector to the seesaw sector, Seesaw Reheating predicts a richer thermal history, characterized by an intermediate relativistic phase and a subsequent relativistic-to-non-relativistic transition of the mediator before radiation domination. This modified expansion history may leave observable imprints on inflationary observables and on the stochastic gravitational-wave background. Together with its implications for leptogenesis and sterile-neutrino dark matter, these signatures will be explored in forthcoming companion papers.

\section{Acknowledgments}

We thank the authors of Ref.~\cite{Garcia:2026ulw} for drawing our attention to the importance of a statistical treatment of the average Lorentz factor $\langle\gamma_S\rangle$ during the reheating process.
This project has received funding from the European Union's Horizon Europe research and innovation programme under the Marie Skłodowska-Curie Staff Exchange grant agreement No.~101086085 (ASYMMETRY), as well as from the CNRS International Research Project (IRP) UCMN.
We also acknowledge the hospitality and support of Institut Pascal at Université Paris-Saclay during the Paris-Saclay Astroparticle Symposium 2025.

\bibliographystyle{apsrev4-2}
\bibliography{Seesawreheating}

\end{document}